\documentclass[range]{ar2e}

\usepackage{graphicx}
\usepackage{dcolumn}
\usepackage{subfigure}
\usepackage{enumerate}
\graphicspath{{ps}}

\def \asy#1#2{^{+#1}_{-#2}}
\newcommand{\invf}{{\rm fb}^{-1}}
\newcommand{\Bstar}{B^{*}}

\newcommand{\nubar}{\bar{\nu}}
\newcommand{\mumu}{\mu^+\mu^-}
\newcommand{\tautau}{\tau^+\tau^-}
\newcommand{\dilep}{\ell^+\ell^-}
\newcommand{\invis}{\;{\mathrm{invisible}}}
\newcommand{\recoil}{_{\mathrm{recoil}}}
\newcommand{\rec}{_{\mathrm{rec}}}

\newcommand{\lmh}{A^0}

\newcommand{\emu}{e^+\mu^-}

\newcommand{\emfo}{\times 10^{-4}}
\newcommand{\emfi}{\times 10^{-5}}
\newcommand{\emsi}{\times 10^{-6}}

\newcommand{\efo}{\times 10^{4}}

\newcommand{\esi}{\times 10^{6}}

\def \etalbelle{{\it et\,al.} (Belle Collab.) }
\def \etalcleo{{\it et\,al.} (CLEO Collab.) }
\def \etalbabar{{\it et\,al.} (BaBar Collab.) }
\def \etalcdf{{\it et\,al.} (CDF Collab.) }
\def \etalatlas{{\it et\,al.} (ATLAS Collab.) }
\def \etald0{{\it et\,al.} (D0 Collab.) }

\def \etalhpqcd{{\it et\,al.} (HPQCD Collab.) }
\def \etalcms{{\it et\,al.} (CMS Collab.) }
\def \etalcusb{{\it et\,al.} (CUSB Collab.) }
\def \etale835{{\it et\,al.} (Fermilab E835 Collab.) }

\def \etalqwg{{\it et\,al.} (Quarkonium Working Group) }

\def \etallhcb{{\it et\,al.} (LHCb Collab.)}

\newcommand{\babar}{{\sc BaBar }}

\newcommand{\kev}{\,\mathrm{ keV}}
\newcommand{\mev}{\,\mathrm{ MeV}}

\newcommand{\mevm}{\mathrm{ MeV}/c^2}
\newcommand{\gev}{\,\mathrm{ GeV}}

\newcommand{\gevm}{\mathrm{ GeV}/c^2}

\newcommand{\ee}{e^+e^-}

\newcommand{\hb}{h_b({\mathrm{1P}})}
\newcommand{\hc}{h_c({\mathrm{1P}})}
\newcommand{\hbp}{h_b({\mathrm{2P}})}
\newcommand{\hbn}{h_b({\mathrm{nP}})}
\newcommand{\hn}{h({\mathrm{nP}})}
\newcommand{\etab}{\eta_b({\mathrm{1S}})}
\newcommand{\etabp}{\eta_b({\mathrm{2S}})}
\newcommand{\etabn}{\eta_b({\mathrm{nS}})}

\newcommand{\psip}{\psi'}

\newcommand{\pip}{\pi^{+}}

\newcommand{\pim}{\pi^{-}}

\newcommand{\br}{\mathcal{B}}

\newcommand{\etal}{{\em et al. }}

\newcommand{\goesto}{\rightarrow}
\newcommand{\bbbar}{\mbox{$b\bar{b}$}}

\newcommand{\jpsi}{\mbox{$ J/\psi$}}
\newcommand{\jpc}{\mbox{$J^{PC}$}}
\newcommand{\omm}{\mbox{$1^{--}$}}
\newcommand{\piz}{\mbox{$\pi$}^{0}}
\newcommand{\pipi}{\mbox{$\pi$}^{+}\mbox{$\pi$}^{-}}
\newcommand{\pizpiz}{\mbox{$\pi$}^{0}\mbox{$\pi$}^{0}}
\newcommand{\upsid}{\mbox{$\Upsilon$}{\rm (1D)}}
\newcommand{\upskd}{\mbox{$\Upsilon$}{\rm (kD)}}
\newcommand{\upsi}{\mbox{$\Upsilon$}{\rm (1S)}}
\newcommand{\upsii}{\mbox{$\Upsilon$}{\rm (2S)}}
\newcommand{\upsiii}{\mbox{$\Upsilon$}{\rm (3S)}}
\newcommand{\upsiv}{\mbox{$\Upsilon$}{\rm (4S)}}
\newcommand{\upsv}{\mbox{$\Upsilon$}{\rm (5S)}}
\newcommand{\upsns}{\mbox{$\Upsilon$}{\rm (nS)}}
\newcommand{\upsms}{\mbox{$\Upsilon$}{\rm (mS)}}
\newcommand{\chibppj}{\mbox{$\chi_{bJ}({\mathrm{3P}})$}}
\newcommand{\chibpj}{\mbox{$\chi_{bJ}({\mathrm{2P}})$}}
\newcommand{\chibj}{\mbox{$\chi_{bJ}({\mathrm{1P}})$}}
\newcommand{\chibjone}{\mbox{$\chi_{bJ}({\mathrm{1P}})$}}

\newcommand{\chibone}{\mbox{$\chi_{b1}({\mathrm{1P}})$}}
\newcommand{\chibtwo}{\mbox{$\chi_{b2}({\mathrm{1P}})$}}

\newcommand{\chibpone}{\mbox{$\chi_{b1}({\mathrm{2P}})$}}
\newcommand{\chibptwo}{\mbox{$\chi_{b2}({\mathrm{2P}})$}}
\newcommand{\chibppzero}{\mbox{$\chi_{b0}({\mathrm{3P}})$}}
\newcommand{\chibppone}{\mbox{$\chi_{b1}({\mathrm{3P}})$}}
\newcommand{\chibpptwo}{\mbox{$\chi_{b2}({\mathrm{3P}})$}}
\newcommand{\chinp}{\mbox{$\chi_{J}({\mathrm{nP}})$}}
\newcommand{\chibnp}{\mbox{$\chi_{bJ}({\mathrm{nP}})$}}
\newcommand{\chibmp}{\mbox{$\chi_{b}({\mathrm{mP}})$}}

\newcommand{\bea}{\begin{eqnarray}}
\newcommand{\beq}{\begin{equation}}
\newcommand{\eea}{\end{eqnarray}}
\newcommand{\eeq}{\end{equation}}

\newcommand{\osz}{^1{\mathrm{S}}_0}
\newcommand{\tso}{^3{\mathrm{S}}_1}
\newcommand{\opo}{^1{\mathrm{P}}_1}
\newcommand{\hbot}{h_b({\mathrm{1P,2P}})}
\newcommand{\chibjot}{\chi_{bJ}({\mathrm{1P,2P}})}
\newcommand{\chibpot}{\chi_{b1,2}({\mathrm{2P}})}
\newcommand{\upsivv}{\Upsilon({\mathrm{4S,5S}})}
\newcommand{\upsot}{\Upsilon({\mathrm{1S,2S}})}
\newcommand{\upstth}{\Upsilon({\mathrm{2S,3S}})}
\newcommand{\upsto}{\Upsilon({\mathrm{2S,1S}})}
\newcommand{\upstht}{\Upsilon({\mathrm{3S,2S}})}
\newcommand{\upsott}{\Upsilon({\mathrm{1S,2S,3S}})}
\newcommand{\upsotd}{\Upsilon({\mathrm{1S,2S,1D}})}
\newcommand{\etabot}{\eta_b({\mathrm{1S,2S}})}
\newcommand{\upsidtwo}{\Upsilon({1^3{\mathrm{D}}_2})}
\newcommand{\tpj}{^3{\mathrm{P}}_J}

\begin{document}

\jname{Annual Review of Nuclear and Particle Science}
\jyear{2013}
\jvol{63}
\ARinfo{1056-8700/97/0610-00}

\title{Recent Results in Bottomonium}

\markboth{Patrignani, Pedlar, and Rosner}{Recent Results in Bottomonium}

\author{C. Patrignani
\affiliation{Dipartimento di Fisica, Universit\`a\ di Genova and I.N.F.N. Sezione di Genova}
T. K. Pedlar  
\affiliation{Department of Physics, Luther College}
J. L. Rosner
\affiliation{Enrico Fermi Institute and Department of Physics, University of
Chicago}}


\begin{abstract}
Great strides have been made in the understanding of bound states of a
bottom quark $b$ and its antiquark $\bar b$ since the discovery of the
first $\Upsilon$ resonances in 1977.  These {\it bottomonium} bound states
have a rich spectrum whose masses and transition amplitudes shed valuable
light on the strong interactions.  The present article reviews
some recent developments in bottomonium physics.  These include the
discovery of the spin-singlet states $\etabot$ and $\hbot$,
the first D-wave states, one or more candidates for spin-triplet $\chibppj$
excitations, and above-threshold states with strong transitions to states below
threshold.  Information on transitions, production, and signatures
of new physics is also presented.
\end{abstract}

\maketitle

\section{INTRODUCTION}

\subsection{$\Upsilon({\mathrm{1S, 2S, 3S?}})$ discovery and similarity to
charmonium}

The first evidence for {\it bottomonium}, the bound states of a bottom quark
$b$ and the corresponding antiquark $\bar b$, was seen in the spectrum of
$\mu^+ \mu^-$ pairs produced in 400 GeV proton-nucleus collisions at
Fermilab \cite{Herb77,Innes77}.  Evidence was presented for the $\upsi$
and $\upsii$ at 9.46 and 10.02 GeV/$c^2$, with a hint of a higher-mass
state now understood to be the $\upsiii$ at 10.35 GeV/$c^2$.

The spacing between 1S, 2S, and 3S levels of bottomonium resembles
that in charmonium, suggesting a simple inter-quark potential $V(r) = C \log
(r/r_0)$ \cite{Quigg77,Quigg79} interpolating between the short-distance $1/r$
and long-distance $r$ behaviors expected in quantum chromodynamics (QCD)
\cite{Eichten75,Eichten78,Eichten80}.  Many successful predictions of
bottomonium properties followed from such QCD-inspired potentials.

\subsection{$\upsns$ are bound states of $b$ and $\bar{b}$}

The $\upsns$ states were immediate candidates for members of a new system of
quark-antiquark bound pairs, but the charge of the new quark was not yet
established.  Several years earlier Kobayashi and Maskawa \cite{Kobayashi73}
had proposed a six-quark model with doublets $(u,d)$, $(c,s)$, and $(t,b)$
to explain CP violation.  The $u$ (up), $d$ (down), $c$ (charm), and $s$
(strange) quarks were already known in 1977, but the new quark could be either
$t$ (top) or $b$ bottom.  The measurement of the decay rates into lepton pairs
of the two lowest-lying states $\upsi$ and $\upsii$ at the
electron-positron collider DORIS in Germany \cite{Darden78} established the
magnitude of the charge of the new quark as 1/3 \cite{Rosner78}, solidifying
its role as the $b$. It took 17 more years for the top quark to
be identified at the Fermilab Tevatron \cite{Abe95,Abachi95}.

\subsection{Major players: Fermilab, DORIS, Cornell, Belle, {\sc BaBar},
$\ldots$}

A number of laboratories have played a role in the study of bottomonium.
Following the discoveries in 1977 at the Fermilab fixed-target experiment
E288 \cite{Herb77,Innes77}, the electron-positron colliders DORIS (at
DESY in Hamburg) and CESR (at Cornell University) made key measurements
of the properties of $b \bar b$ states.  For two reviews of the early history
of these measurements see Refs.\ \cite{Kwong87} and \cite{Buchmuller88}; for a
history of CESR and its main particle detector CLEO see \cite{Berkelman04}.
Subsequently major contributions were made by experiments at the asymmetric
electron-positron colliders PEP-II ({\sc BaBar} Detector) and KEK-B (Belle
Detector), the CDF and D0 detectors at the Fermilab Tevatron
proton-antiproton collider, and the ATLAS, CMS, and LHCb detectors at the CERN
Large Hadron Collider (LHC).  One review of the second two decades of
bottomonium spectroscopy is Ref.\ \cite{Eichten08}.  For a more recent and
comprehensive survey of quarkonium results see Ref.\ \cite{Brambilla11}.

\begin{figure}[h]
\begin{center}
\includegraphics[width = 0.8\textwidth]{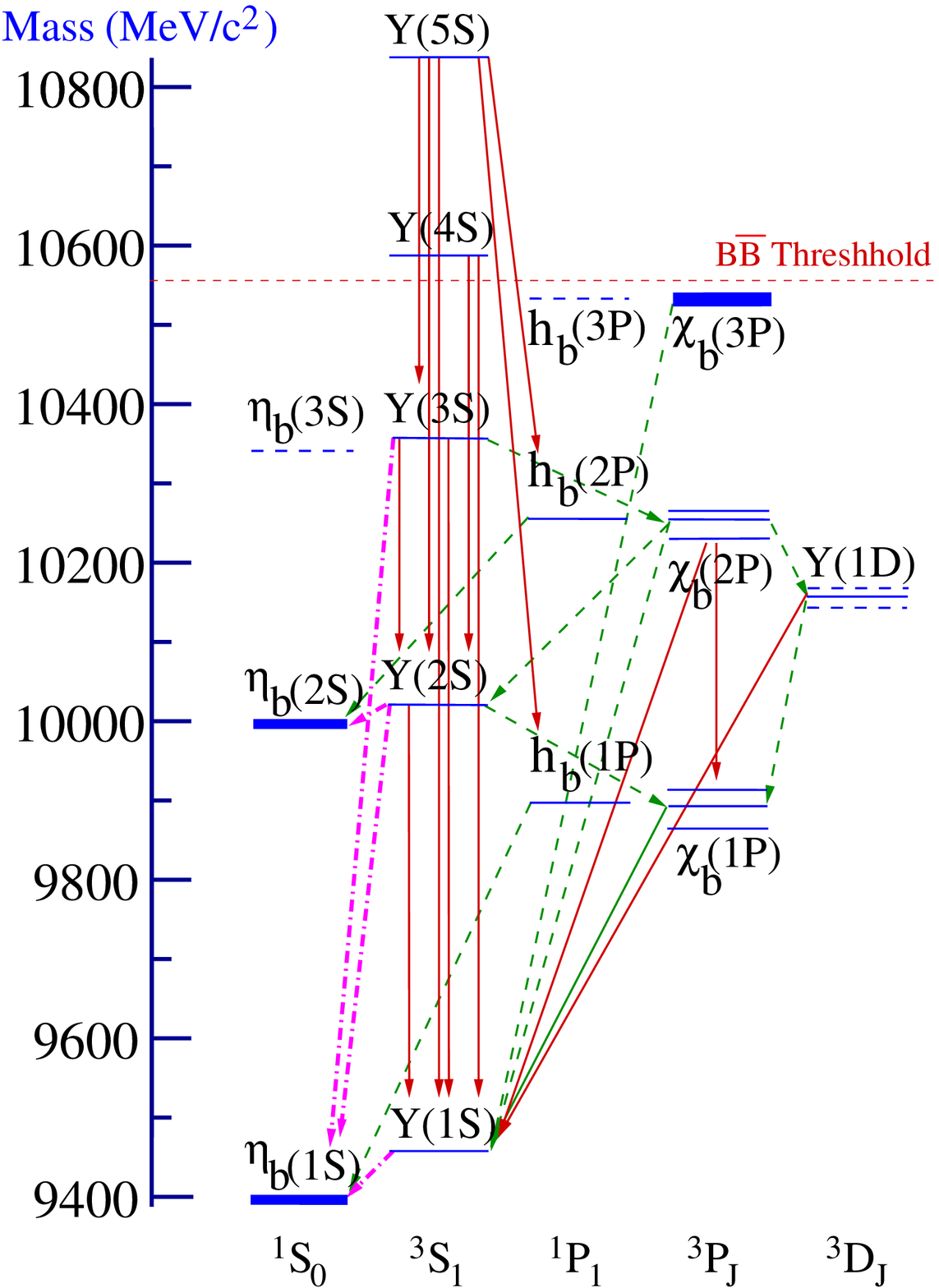}
\end{center}
\caption{Current knowledge of the bottomonium system.  Solid lines correspond
to known states while dashed lines are predicted ones.  The thicker lines
indicate the range of measured masses for newly discovered states.
(Solid, dashed, dot-dashed) arrows denote (hadronic, electric dipole [E1],
and magnetic dipole [M1]) transitions, respectively.
\label{fig:ups}}
\end{figure}

\section{SPECTROSCOPY}

\subsection{Overview:  masses, spins, and parities of known states}

In Fig.\ \ref{fig:ups} we display the current state of knowledge about
bottomonium states and transitions among them.  Such states must
decay either via $b \bar b$ annihilation or through transitions to lower
bottomonium states, with typical total widths in the tens to hundreds of keV.
States above flavor threshold can decay to pairs of flavored mesons and have
typical total widths larger by at least two orders of magnitude.

Quark-antiquark composites with total spin $S=0,1$ and orbital angular momentum
$L$ have parity and charge-conjugation eigenvalues $P = (-1)^{L+1}$ and
$C = (-1)^{L+S}$.  We use the notation S, P, D, $\ldots$ for states with
$L=0,~1,~2,\ldots$, with a superscript prefix denoting the spin multiplicity
$2S+1$ and a subscript denoting total angular momentum $J$.  An integer
preceding the angular momentum letter denotes the radial quantum number
$n = 1, 2, \ldots$.  Thus, the full notation for the $\upsi$ is $1\tso$.
$\upsns$ states are easily produced in hadronic interactions and through
virtual photons in $e^+ e^-$ collisions.

Before reviewing discoveries of the past few years (in roughly chronological
order), we discuss briefly the notation in Fig.\ \ref{fig:ups}.
The spin-singlet partners ($\osz$) of the $\upsns$ are called
$\etabn$.  The first of these was discovered only a few years ago
\cite{Aubert08,Bonvicini10} and its 2S partner within the past year
\cite{Mizuk12}.

Quarkonium $\tpj$ states are denoted by $\chinp$; those of bottomonium
are called $\chibnp$. The 1P and 2P states were discovered \cite{Han82,%
Klopfenstein83} in experiments by the CUSB (Columbia University -- Stony Brook)
Collaboration at the CESR $e^+ e^-$ collider.  They may be produced through
electric dipole radiative transitions from excited $\upsns$ states; in turn,
they decay via electric dipole transitions to lower $\upsns$ states.  They may
also be produced directly in hadronic interactions, as evidenced by the recent
observation of the $\chibppj$ states in hadronic collisions at the LHC
\cite{Aad12} and the Tevatron \cite{Abazov12} and decaying to $\gamma +\upsot$.

Quarkonium $\opo$ states are called $\hn$.  The charmonium state $\hc$
was the last $c \bar c$ state to be discovered below charmed meson pair
threshold \cite{Rosner05,Rubin05,Andreotti05}.  The corresponding bottomonium
state $\hb$ and the excited state $\hbp$ were seen relatively
recently \cite{Adachi12}.

The states in Fig.\ \ref{fig:ups} denoted $\Upsilon({\mathrm{n}}^3{\mathrm{D}})$ are spin-triplets
with orbital angular momentum $L=2$ and thus have $J=1,2,3$.  Evidence for them
will now be discussed.

\subsection{D-wave states(s):  CUSB, CLEO, {\sc BaBar}}

Quarkonium potential models (e.g., \cite{Kwong88})
predict masses of $\upsid$ and $\Upsilon(2{\mathrm{D}})$ states within a
narrow range around 10.16 and 10.44 MeV \cite{Godfrey01b}.  An early search
for transitions $\upsiii \to \gamma_1 \chibpj \to \gamma_1 \gamma_2
\upsid \to \gamma_1 \gamma_2 \gamma_3 \chibj [\to \gamma_1 \gamma_2
\gamma_3 \gamma_4 \upsi]$ by the CUSB Collaboration found no
evidence for the $\upsid$ states \cite{Heintz92}.  However, with
larger statistics and an excellent CsI electromagnetic calorimeter, the CLEO
Collaboration was able to identify  
events with four photons corresponding to the above cascades \cite{Bonvicini04}.

The dominant contribution to this chain is predicted to come from
$\upsidtwo$ \cite{Kwong88,Godfrey01b}.  Under this assumption, the mass
of the new state is $M[\upsidtwo] = (10.161.1 \pm 0.6 \pm 1.6)$ MeV.
The mass is consistent with potential model calculations, as is the product
branching fraction (assuming $J=2$) for the above four transitions followed by
$\upsi \to e^+e^-$ or $\mu^+ \mu^-$: ${\cal B} = (2.5 \pm 0.5 \pm 0.5)
\times 10^{-5}$.  More recently, the {\sc BaBar} Collaboration \cite{PdAS10}
has measured the mass of the $\upsidtwo$ state
using its decay to $\pi^+ \pi^- \upsi$ to be $M[\upsidtwo =
(10164.5 \pm 0.8 \pm 0.5)$ MeV/$c^2$, also finding 
weak evidence for the
$J=1$ and $J=3$ states.  The angular distributions are consistent with
expectations for the $J=2$ state. 

\subsection{$\etab$ discovery; splitting from $\upsi$}

The ground state of bottomonium, $\etab$, was predicted by a variety of
QCD-inspired potential models to lie between 35 and 100 MeV below the
$\upsi$ \cite{Godfrey01a}.  The large range of predictions arose to a great
extent from uncertainties in relativistic corrections and in evaluation of
the square of the wave function at the origin, entering into the prediction
of the hyperfine splitting. 

The (allowed)
magnetic dipole (M1) transition $\upsi \to \gamma \etab$ leads to a very
soft photon which is difficult to distinguish from the many photons due
to $\upsi \to \pi^0 X \to \gamma \gamma X$.  This led to the attempt to
observe the much more energetic photon in $\upsiii \to \gamma \etab$ (911 MeV
for $M[\etab] = 9.4~\gevm$).  The CLEO Collaboration searched for
for this transition in 5.9 million $\upsiii$ decays, finding only an upper
limit on the rate \cite{Artuso05}.  With $(109 \pm 1)$
million $\upsiii$ decays, the {\sc BaBar} Collaboration observed a state lying
$71^{+3.1}_{-2.3} ({\rm stat}) \pm 2.7({\rm syst})$ MeV below the $\upsi$,
while observation of the transition $\upsii \to \gamma \etab$ led to a
hyperfine splitting of $(66.1^{+4.9}_{-4.8}\pm 2.0)$ MeV.

The {\sc BaBar} discovery employed subtraction of substantial backgrounds
at lower photon energies.  At the $\upsiii$, initial state radiation (ISR) to
$\upsi$ results in a photon energy of 856 MeV, while radiative transitions to
the $\chibpj$ states followed by $\chibpj \to \gamma \upsi$ lead to a
broad cluster of Doppler-broadened lines around 764 MeV.  Using this technique
the CLEO Collaboration was able to identify the $\upsiii \to \gamma \upsi$
transition photon \cite{Bonvicini10}, measuring the hyperfine splitting from
$\upsi$ to be $(68.5 \pm 6.6 \pm 2.0)$ MeV.  A more recent measurement of
$M[\etab]$ by the Belle Collaboration using a different technique (described
below) gives a slightly smaller hyperfine splitting of $(57.9 \pm 1.5 \pm1.8)$ 
MeV \cite{Mizuk12}, in agreement with the most recent calculations based on
lattice QCD \cite{Meinel10,Dowdall12}.

\subsection{Discovery of $\hb$, $\hbp$, and $\etabp$\label{sec:singlets}}

The spin-singlet P-wave states of quarkonium are expected to lie very
close in mass to the spin-weighted average of the triplet states, as the
hyperfine splitting in leading non-relativistic order is proportional to
the square of the wave function at the origin, which vanishes for P-wave
states.  Such was found to be the case for charmonium \cite{Rosner05,Rubin05,%
Andreotti05,Dobbs08}.

Evidence for the lowest $^1P_1$ state of bottomonium, $\hb$, was first
noted by the {\sc BaBar} Collaboration \cite{Lees11} in the transition
$\upsiii \to \piz\hb \to \piz \gamma \etab$.  The first significant
signal for this state, however, came from an unexpected quarter:
hadronic transitions to $\hbn$ from the $\upsv$, a bottomonium state lying 
well above the open flavor threshold. 

Two observations prompted the search for $\hbn$ in such an unlikely process.  
The CLEO Collaboration had observed the production of $h_c$ in $e^+ e^-
\to \pi^+ \pi^- h_c$ at a center-of-mass energy of 4.16 GeV, lying
above charm threshold, at a rate comparable to that for $e^+ e^- \to \pi^+
\pi^- J/\psi$ \cite{Pedlar11}.  This was a surprise as it indicated none of
the expected suppression for a $c$-quark spin-flip process.  Meanwhile, the
Belle Collaboration had observed anomalously high rates for $e^+ e^- 
\goesto \pi^+ \pi^- \upsns$ at energies above bottom-meson pair production
threshold (the dashed horizontal line in Fig.\ \ref{fig:ups})~\cite{Chen08}.
These curious enhancements stimulated a search for $e^+ e^- \to \pi^+ \pi^-
h_b$ near the $\upsv$ ($M = 10865$ MeV/$c^2$) resonance \cite{Adachi12}.

\begin{figure}[htp]
\begin{center}
\includegraphics[width=5in]{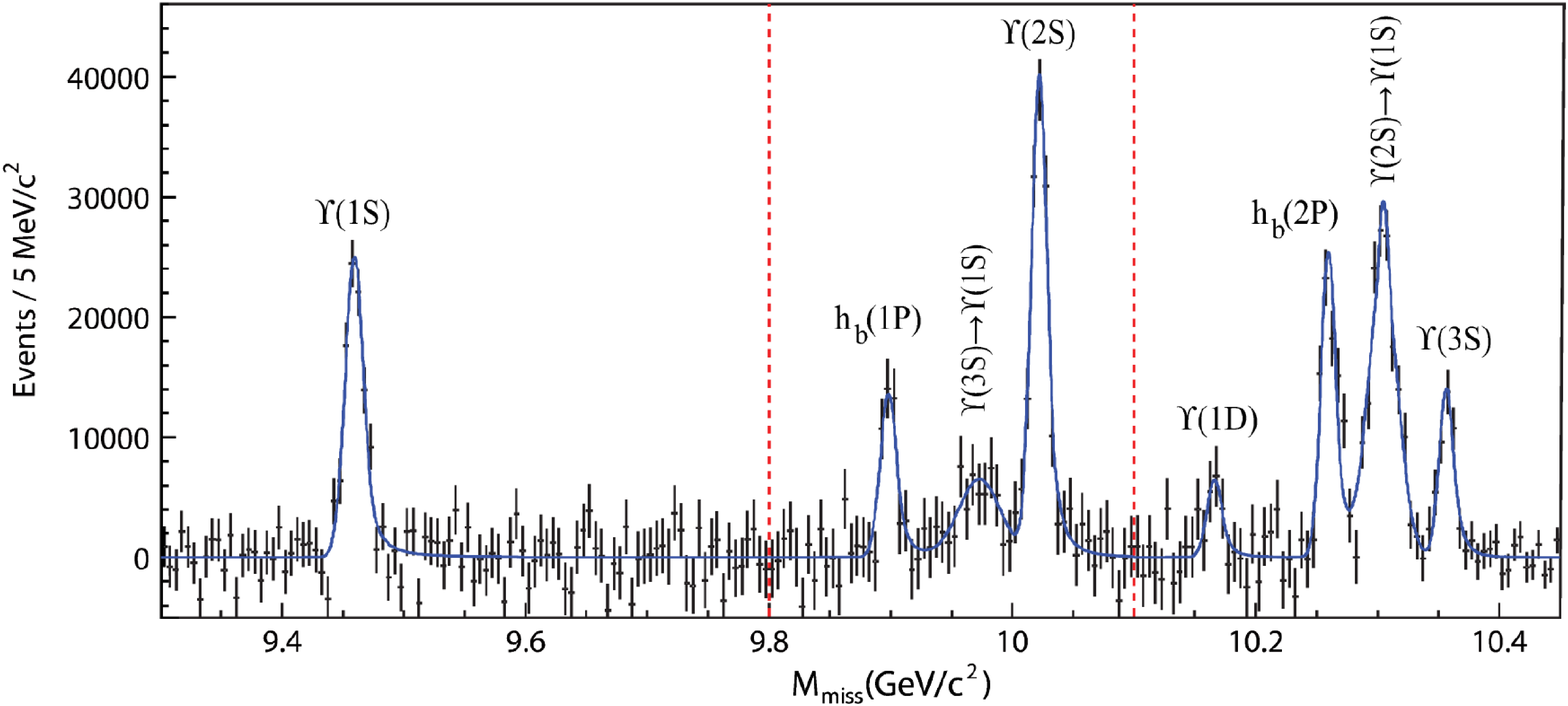}
\caption{Recoil mass spectrum against The $\pipi$ recoil mass spectrum from Belle~\cite{Adachi12}.  
In the figure, the peaks in the recoil mass spectrum arise from $\upsv\goesto\pipi\upsns$ with
$n=1,2,3$, $(\upsiii,\upsii)\to\pipi\upsi$ arising from $\upsiii$ and $\upsii$ produced via ISR, 
and direct $\upsv\to\pipi(\hb,\hbp)$. The peak near 10.16 $\gevm$ constitutes 
evidence for $\upsv\to\pipi\upsid$.~\label{mmpipi_belle}}   
\end{center}
\end{figure}

The search succeeded beyond expectations.  Not only was the $\hb$ seen, very
close to the expected spin-weighted average $\overline{M}[\chibj]$, but
the $\hbp$ was observed with even greater significance, near $\overline{M}
[\chibpj]$ (see Fig.\ \ref{fig:ups}).  Belle observed these states in the
recoil mass spectrum against $\pipi$.  In Fig.~\ref{mmpipi_belle} are shown
the Belle observations of dipion transitions from
$\upsv$ to all three sub-threshold $\upsns$ states, dipion transitions
$(\upstth\to\pipi\upsi$ from $\upstth$ produced via ISR, 
direct $\pipi$ transitions to $\hbot$ and evidence for a $\pipi$ 
transition to $\upsid$.  These results were updated in Ref.\
\cite{Mizuk12}, leading to $M[\hb] = (9899.1\pm0.4\pm 1.0)$ MeV/$c^2$, which is
$(0.8 \pm 1.1)$ MeV/$c^2$ below $\overline{M}[\chibj]$, and $M[\hbp] =
(10259.8 \pm 0.5 \pm 1.1)$ MeV/$c^2$, which is $(0.5 \pm 1.2)$ MeV/$c^2$ below
$\overline{M}[\chibpj]$.  Thus, no evidence was found for hyperfine
splittings in the P-wave bottomonium states.

The Belle Collaboration also was able to observe the transitions $\hb \to
\gamma \etab$, $\hbp \to \gamma \etab$, and $\hbp \to \gamma 
\etabp$~\cite{Mizuk12}. With this
they discovered a new state, $\etabp$ with 
$M[\etabp] = (9999.0 \pm 3.5^{+2.8}_{-1.9})$
MeV/$c^2$, and made the world's most precise measurement of $M[\etab] = (9402.4
\pm 1.5 \pm 1.8)$ MeV/$c^2$.  Additional evidence for the $\etabp$ was obtained
in a study \cite{Dobbs12a} using CLEO data, resulting in a substantially lower 
mass $M[\etabp] = (9974.6\pm 2.3 \pm 2.1)\mevm$, and, in addition, a mass
measurement of $\etab$: $M[\etab] = (9393.2\pm 2.3 \pm 2.1)\mevm$
based on $3 \sigma$ evidence for $\Upsilon(1S) \to \gamma \eta_b(1S)$.

\subsection{The $\chibppj$ states}

Some bottomonium states, though expected to lie below $B \bar B$ flavor
threshold and thus expected to be narrow, are high enough in mass that they
cannot be reached by one or more electric dipole transitions from a
higher-lying $\upsns$ state also lying below flavor threshold.  They are
thus difficult to observe in $e^+ e^-$ collisions.  The $\chibppj$ states
are one example.  Most bottomonium potential models predict them to lie
below flavor threshold; a representative set of masses \cite{Kwong88} is
$M[\chibppzero] = 10500.7$ MeV/$c^2$, $M[\chibppone] = 10516.0$ MeV/$c^2$,
and $M[\chibpptwo] = 10526.4$ MeV/$c^2$, to be compared with $2 M(B) \simeq
10560$ MeV/$c^2$.  Other models predict values within ten or twenty MeV/$c^2$
of these [see, e.g., \cite{Daghigian87,Motyka98}].

The $\chibppj$ states can be produced in hadronic interactions, for example
through two-gluon fusion (see Sec.\ 4.3), and then can be observed through
their electric dipole transitions to lower $\upsns$ states.  In this
manner the ATLAS Collaboration \cite{Aad12} reported observing not only the
previously known decays $\chibj \to \gamma \upsi$ and $\chibpj
\to \gamma \upsot$, but also $\chibppj \to \gamma \upsot$.

Using converted photons (with superior energy resolution), the ATLAS
Collaboration measures the weighted average of the masses contributing
to these last two transitions to be $\bar m_3 = 10530 \pm 5 \pm 9$ MeV.
It is most likely that this receives little contribution from the
$\chibppzero$, whose partial width to hadronic states is expected to be
considerably larger than that of the other two states \cite{Kwong88}, so
it is probably to be compared with some weighted combination of the predicted
masses of $\chibppone$ and $\chibpptwo$.

The D0 Collaboration  also has observed a similar structure,
using converted photons~\cite{Abazov12}.  
Calibrating their energy using the known transitions
involving the $\chibjot$ states, they find $\bar m_3 = 10551\pm 14\pm 17$
MeV/$c^2$.

\subsection{Above-threshold (molecular?) states}

For years, a puzzle in the quark model was why all the observed hadrons
appeared to be composed either of quark-antiquark (mesons) or three quarks
(baryons).  These configurations can exist in color singlets, but they
are not the only ones.  What about tetraquarks ($q q \bar q \bar q$),
pentaquarks ($qqqq \bar q$) or more complicated objects?  After all, nuclei
do exist, though principally they are thought of as clusters of baryons.

Since the earliest days of hadron spectroscopy, it was recognized that
properties of levels can be strongly influenced by nearby channels.  An
early example was the resonance $\Lambda(1405)$, decaying to $\Sigma \pi$
but strongly influenced by the $\bar K N$ threshold lying just above it
\cite{Dalitz59}.  Another example is the $f_0(980)$, decaying to $\pi \pi$
but behaving remarkably like a $K \bar K$ bound state or ``molecule''
\cite{Flatte72}.  The description of such states requires one to go beyond
a simple $qqq$ or $q \bar q$ picture.

Molecular ``charmonium'' was proposed soon after the discovery of charm
\cite{Voloshin76,DeRujula77}.  The attraction of a charmed meson $D = c \bar q$
and an anticharmed meson $\bar D = \bar c q'$ could be viewed as a molecular
force mediated by the light quarks with the $c$ and $\bar c$ acting as
spectators.  An excellent candidate for such a state was identified in 2003
by the Belle Collaboration \cite{Choi03}: the X(3872), originally seen
decaying to $\pi^+ \pi^- J/\psi$.  [For a detailed description of this and
related states see Ref.\ \cite{Godfrey08}.]  Its mass is very close to
$M(D^0) + M(D^{*0})$.  The consensus is that it shares many properties of
a $D^0$--$\bar D^{*0}$ molecule, with some admixture of a $^3P_1$ charmonium
state to explain its observed decay to $\gamma J/\psi$. 

It now appears that there are candidates for molecular bottomonium as well.
The surprisingly large rates for transitions from $\Upsilon({\mathrm{5S,10865}})$ to
$\pi^+ \pi^- \upsott$ and $\pi^+ \pi^- \hbot$ appear largely to be due to a
sequential process in which a pion and a $B \bar B^*$, $B^* \bar B$, or $B^*
\bar B^*$ are first produced, forming resonances which then emit another pion
to reach the final state \cite{Adachi11}.  These bottomonium-related states
are electrically charged, and have come to be called $Z_b(10610)$ and
$Z_b(10650)$.

Their large mass, indicating the presence of two bottom quarks in their
composition, together with their possession of electrical charge, 
marks these states as necessarily unconventional. Their
proximity to the $B\Bstar$ and $\Bstar\Bstar$ thresholds
at $M(B) + M(B^*) = 10604.6$ MeV/$c^2$ and $2M(B^*) = 10650.2$ MeV/$c^2$
makes their identification as molecular states an attractive possiblity
\cite{Bondar11}.  Their parity is positive as would be expected for S-wave
molecular states.  Their participation in the intermediate state of the
transitions from the $\upsv$ which produce $\hbot$ as strongly as
$\upsott$ follows naturally from a careful accounting of the
ratio of $b \bar b$ triplet and singlet in their wave functions.  The
molecular description of these states predicts equal total widths (as
observed) and several other decay modes such as $\etab \rho$.
The properties of these states are summarized in Table \ref{tab:Zs}.  Both 
have isospin $I=1$, positive G-parity, and are determined to have spin-parity
$J^P = 1^+$ by angular analysis of their production and decay kinematics.  

On the basis of these properties, one might additionally expect the existence
of neutral partners to these charged
$Z_b$ states. The Belle Collaboration has recently 
announced evidence at a significance of $4.9\sigma$ 
for such a state at a mass of $10609\asy{8}{6}\pm 6~\mevm$ 
in $\pizpiz$ transitions from $\upsv$~\cite{Adachi12a}.  
The evidence for a neutral state corresponding to the $Z_b(10650)$ in the same
analysis is insignificant. 

Finally, the Belle Collaboration has recently announced the results of their
investigation of three-body decays of $\upsv$ to final states 
of $[B^{(*)}B^{(*)}]^{\pm}\pi^{\mp}$~\cite{Adachi12b}.  In this study it was
observed that the $Z_b(10610)$ decays dominantly $(86.0\pm 3.6)\%$ of the time
to $BB^*$, while the $Z_b(10650)$ decays dominantly $(73.4\pm 7.0)\%$ to
$B^*B^*$, which serves as additional evidence in favor of 
the molecular interpretations of the charged $Z_b$ states. 

\begin{table}
\caption{Properties of candidates for molecular bottomonium.
\label{tab:Zs}}
\begin{center}
\begin{tabular}{c c c} \hline \hline
State      & Mass     & Width \\
       & (MeV/$c^2$)  & (MeV)\\ \hline
$Z_b(10610)$ & $10608.4\pm2.0$ & $15.6\pm2.5$ \\
$Z_b(10650)$ & $10653.2\pm1.5$ & $14.4\pm3.2$ \\ \hline \hline
\end{tabular}
\end{center}
\end{table}

\subsection{The $Y_b(10890)$}

As noted in Sec.~\ref{sec:singlets}, 
in 2008, the Belle Collaboration observed anomalously large rates (up to
two orders of magnitude larger than expected) for the processes 
$\upsv\goesto\pipi\upsot$ if the $\upsv$ is interpreted as the fourth
radial excitation of the $\omm$ $\upsi$ state~\cite{Chen08}.
(See discussion later in Sec.~\ref{sec:ups5shadronic}.)
This led them to conduct a careful scan of the cross section for 
$\pipi\upsns$ in the vicinity of the known mass of the $\upsv$, 10865 $\gevm$,
taking data at center of mass energies between 10.83 and 11.02 $\gev$.  
In this study, they observed a peak in $\sigma(e^+e^- \to
\upsns \pi^+ \pi^-)$ $(n=1,2,3)$ at an energy of $(10888^{+2.7}_{-2.6}
\pm 1.2)$ MeV with a width of $(30.7^{+8.3}_{-7.0} \pm 3.1)$ MeV
\cite{Adachi10}, as will be discussed in more detail in 
Sec.~\ref{sec:ups5shadronic}.
The {\sc BaBar} Collaboration also recently measured two peaks in the $e^+ e^-
\to b \bar b$ cross section, one at $(10876 \pm 2)$ MeV with width $(43 \pm 4)$
MeV and the other at $(10996 \pm 2)$ MeV with width $(37 \pm 3)$ MeV
\cite{Aubert09}.  

 The displacement of the peak measured in $\upsns \pi^+ \pi^-$ by Belle and the lower {\sc BaBar} peak measured in $b\bar b$ decays
has led to the suggestion that 
these peaks may not, in fact, be the $\upsv$ but rather some exotic state
\cite{Liu12}.

\subsection{Special aspects of $\upsv$}

Both $B \bar B{^*}$ and $B^* \bar B^*$ can interact by exchanging pions,
while in order to do so $B \bar B$ states have to undergo virtual transitions
to $B^* \bar B^*$.  This may explain the existence of the ``molecular''
states $Z_b(10610)$ and $Z_b(10650)$, with the $\upsiv$ behaving more
as a conventional $b \bar b$ level.  (See, however, the discussion below of
hadronic transitions of $\upsiv$.)

As the $Z_b$ states have isospin $I=1$, they must be produced in $e^+ e^-$
collisions in association with a pion.  The first $\Upsilon$ resonance which
permits this is the $\upsv$, at 10865 MeV.  This could explain why the
beautiful proliferation of states in $\upsv$ decays is not seen at lower
energies.

\subsection{Theoretical successes and shortcomings}

Potential models based on short-distance gluon exchange and a linear
$b \bar b$ confining potential at large distances have reproduced reasonably
well the spin-triplet $nS$, $nP$, and $1D$ bottomonium
spectra below flavor threshold.  Pioneering work in this area
\cite{Eichten75,Eichten78,Eichten80} has been extended to include the effects
of relativistic corrections (see, e.g., \cite{Ebert03}) and the effects of
coupling to open channels (see, e.g., many references in \cite{Brambilla11}).
Potential model predictions for the masses of the $1D$ and $3P$ levels had a
narrow range of ten or twenty MeV and were verified within that range.
For the lowest-lying levels, lattice gauge theories (extended to include
the effects of light-quark pairs) have also provided an effective description
(see, e.g., \cite{Gray05}).

One prediction of spin-dependent effects, shared in many approaches, also
had great success.  This is the small value of the hyperfine splitting in
1P and 2P $b \bar b$ states, as confirmed by the $\hb$ and $\hbp$ masses
\cite{Adachi11}.  This may be traced to the vanishing of the P-wave wave
function at the origin.  The hyperfine splitting in $nS$ states posed more
of a problem for potential models, with a wide range of predictions
\cite{Godfrey01a}.  Recent estimates of $M[\upsi] - M[\etab]$ based on lattice
gauge theory \cite{Meinel10,Dowdall12} have been fairly close to the observed
values.

The situation is quite different for states above flavor threshold.  As
exemplified by the states $Z_b(10610)$ and $Z_b(10650)$,
a $b \bar b$ description of such states does not suffice.  They lie just
a few MeV below $M(B)+M(B^*)$ and $2 M(B^*)$, respectively, suggesting that
these channels play an important role in their makeup.  Indeed, successful
predictions result from regarding these as molecular bottomonium states
\cite{Bondar11}.  The discovery of these unique resonances is simply the most 
recent in a series of observations that underscores the importance of 
nearby thresholds in 
spectroscopy~\cite{Eichten78,Eichten80,Liu12,Byers90,Eichten06,Rosner06}.

\section{TRANSITIONS}

\subsection{Decays of $\upsi$ to light-quark states}

The $\upsi$ is expected to decay mainly [$(81.7 \pm 0.7)\%$] via three gluons,
with $(2.2 \pm 0.6)\%$ to two gluons and a photon \cite{PDG}.  The two- and
three-gluon channels provide an entry to many potential final states, including
states made of pure glue (glueballs), light Higgs bosons, and states made of
lighter quarks.  However, until recently relatively few of these had been
identified, in contrast to the wealth of final states seen in $J/\psi$ decays.
(One interesting branching fraction is ${\cal B}[\upsi \to \eta'(958) + X] =
(2.94 \pm 0.24)\%$ \cite{PDG}.)

This situation changed with the publication of observations by Belle of several
exclusive hadronic decays of $\upsii$ and $\upsi$~\cite{Shen12}, and by CLEO
of seventy-three exclusive hadronic decay modes of $\upsi$ and seventeen of
$\upsii$ \cite{Dobbs12b}.  The branching fractions for these decays ranged from
$0.5 \times 10^{-6}$ to $110 \times 10^{-5}$.  Together with multi-particle
decays of $\chibjot$ to be mentioned below, such results pave the way for a 
more complete understanding of how multi-gluon final states fragment into
hadrons.

\subsection{Decays of $\upsi$ to charm, charmonium}

The mechanism of hadronic quarkonium production still remains obscure
nearly four decades after the discovery of the $J$ in proton-beryllium
collisions.  The inclusive decays of bottomonium states to charmonium can
help to illuminate this mechanism by providing a clean gluonic initial
state.

The CLEO Collaboration studied the processes $\upsi \to J/\psi X,$ $\psi(2S) X$,
$\chibone X$, $\chibtwo X$ using 21 million $\upsi$ decays \cite{Briere04}.
These measurements can test the {\it color octet model} \cite{Braaten95}, in
which a substantial fraction of hadronic quarkonium production proceeds through 
a color-octet component of the quarkonium wave function.  What is found is that
the ratios of the excited charmonium states mentioned above to $J/\psi X$
production are about a factor of two larger than predicted by the color-octet
model~\cite{Cheung96}.
The {\sc BaBar} Collaboration studied the $\upsi \to D^{*\pm}(2010) X$
decays~\cite{Aubert10c} as a function of the scaled $D^{*\pm}$ momentum,
finding the color-octet contribution to be disfavored~\cite{Zhang08}.

\subsection{Multi-particle decays of $\chibjot$}

The decays of heavy quarkonia into states of light-quark hadrons can tell
us how initial parton states consisting
of gluons and quarks turn into observable particles.  Many Monte Carlo
programs make assumptions about this process which have not been fully
tested against data.  The CLEO Collaboration studied the decays of the
$\chibj$ and $\chibpj$ states into 659 different states of
pions, kaons, $\eta$s, and baryons \cite{Asner08}, identifying 14 modes
with significant signals.  The greatest significance was found for signals from
the $J=1$ states.  The selection criteria were limited to those final
states containing 12 or fewer particles, where a $\pi^0 \to \gamma \gamma$
or a $K_S \to \pi^+ \pi^-$ counts for two.  Even with this high multiplicity
it appeared that significant higher multiplicities were being missed.

\subsection{Electric dipole transitions $S \leftrightarrow P \leftrightarrow
D$}

The transitions $\upsns \leftrightarrow \chibmp \leftrightarrow
\upskd$ can occur via single electric dipole photon emission or
absorption.  They occur with observable branching fractions for the narrow
bottomonium states below flavor threshold.  For spin-triplet states, they have
been seen in $\chibj \to \gamma \upsi$, $\upsii \to \gamma
\chibj$, $\upsid \to \gamma \chibj$, $\chibpj \to \gamma
\upsotd$, and $\upsiii \to \gamma \chibjot$.  More
recently they have also been seen in the spin-singlet states:  $\hb \to
\gamma \etab$ and $\hbp \to \gamma \etabot$.  These transitions
satisfy $\Delta L = 1$ with change of parity and no change in quark spin.  They
involve the matrix element of $\vec{r}$ between initial- and final-state wave
functions.  Their rates are given by expressions quoted (e.g.) in
Refs.\ \cite{Eichten80,Kwong88}.  They are illustrated in Fig.\ \ref{fig:ups}.

Transitions $nP \to nS$ or $nD \to nP$ involve initial and final wave functions
with the same number of nodes and have substantial dipole matrix elements.
Although the number of nodes in the wave functions differs for $nP$ and
$[n-1]S$ or $nS$ and $[n-1]P$, the corresponding dipole matrix elements
are also appreciable.  These are the transitions that would be allowed in
the three-dimensional harmonic oscillator.  On the other hand, the transitions
$\upsiii \to \gamma \chibj$ are highly suppressed; in a three-d
harmonic oscillator they would be forbidden.  This
suppression holds for a wide range of power-law potentials \cite{Grant92}.
As a result, dipole matrix elements for these transitions are particularly
sensitive to relativistic corrections, providing a way of discriminating
among various approaches.

\subsection{Forbidden M1:  $\upstth\goesto\gamma\etab$}

The nonrelativistic approximation for magnetic dipole transition rates in
$Q \bar Q$ bound states \cite{Godfrey01a} is
\beq
\Gamma(\tso \to \gamma \osz)=\frac{4}{3}\alpha \frac{e_Q^2}{m_Q^2} I^2 k^2~,
\eeq
where $\alpha$ is the fine-structure constant, $e_Q$ is the quark charge
in units of $|e|$, $m_Q$ is the quark mass, and $k$ is the magnitude of the
photon three-momentum in the rest frame of the decaying particle.  The overlap
integral $I$ is defined by $I = \langle f | j_0(kr/2) | i \rangle$, where
$j_0(kr/2) = \sin(kr/2)/(kr/2) \simeq 1 - (kr)^2/24 + \ldots$.  When the
initial and final states have the same principal quantum numbers, and the
photon energy is low, the overlap integral is close to unity.  However,
when the initial and final principal quantum numbers are not the same, a
{\em forbidden} transition, the leading matrix element
vanishes as a result of the orthogonality of the initial and final wave
functions.  Consequently, the overlap integral becomes proportional to $k^2$.
The $\eta_b$, discussed earlier, was first detected in the forbidden transition
$\upsiii \to\gamma\etab$ by the \babar Collaboration \cite{Aubert08}.
It was then seen by \babar in the transition $\upsii \to\gamma\etab$
\cite{Aubert09aa} and confirmed in $\upsiii \to\gamma\etab$ by CLEO
\cite{Bonvicini10}.  These measurements are compared in Table \ref{tab:M1}.

\begin{table}
\caption{Comparison of measurements of forbidden M1 transitions $\upsns
\to \etab$.
\label{tab:M1}}
\begin{center}
\begin{tabular}{c c c c c} \hline \hline
 & \multicolumn{2}{c}{\babar} & \multicolumn{2}{c}{CLEO \cite{Bonvicini10}} \\
Initial $\Upsilon$ & 2S \cite{Aubert09aa} & 3S \cite{Aubert08} & 2S & 3S \\
\hline
$N(\Upsilon)$ (10$^6$) & 91.6 & $109\pm1$ & $9.32\pm0.19$ & $5.88\pm0.12$ \\
$M[\etab]$ (MeV/$c^2$) & $9394.2^{+4.8}_{-2.3}$$\pm$2.0
 & $9388.9^{+3.1}_{-2.3}$$\pm$2.7 & -- & 9391.8$\pm$6.6$\pm$2.0 \\
${\cal B}[\gamma \etab]$ $(10^{-4})$ & $3.9\pm 1.0^{+1.1}_{-0.9}$
 & 4.8$\pm$0.5$\pm$0.6 (a) & $<8.4$ & $7.1\pm1.8\pm1.3$\\
$\Gamma(\etab$ (MeV/$c^2$) & (b) & 10 (c) & $10\pm5$ (c) & $10\pm5$ (c) \\
\hline \hline
\end{tabular}
\end{center}
\leftline{(a) Updated systematic error in \cite{Aubert09aa}. (b) Consistent
with being dominated by}
\leftline{resolution of 18 MeV/$c^2$.  (c) Assumed.}
\end{table}

The masses measured in these transitions are to be compared with that measured
by Belle \cite{Mizuk12} in the allowed E1 transitions $\hbot \to \gamma
\etab$:  $M[\etab] = (9402.4 \pm 1.5 \pm 1.8)$ MeV/$c^2$.  (This
experiment also measured $\Gamma[\etab] = (10.8^{+4.0+4.5}_{-3.7-2.0})$
MeV/$c^2$.)  This last mass is free from potential distortion of the
spectral shape that occurs in the forbidden M1 transitions due to the
$k^2$ dependence of the overlap integral $I$. (See the discussion concerning
the sensitivity of the determination of mass of the $\eta_c$ produced in 
M1 decays of $\jpsi,\psip$ in Ref.~\cite{Mitchell09}.)  It also agrees better
with recent lattice calculations
\cite{Meinel10,Dowdall12}.  The branching fractions are consistent with some
but not all of the wide range of predictions \cite{Godfrey01a}.

\subsection{Hadronic $\bbbar^{\prime}\goesto \bbbar + X$}

Hadronic transitions of heavy quarkonia provide unique insights into the nature
of hadronization at low momentum transfer, and are generally successfully 
described in terms of multipole moments of the QCD 
field~\cite{Gottfried78,Voloshin79,Yan80}.  Two-pion transitions were the only
hadronic transitions known in the system until the observation by the CLEO
Collaboration of an unexpectedly large rate for the transition $\chibpot)
\goesto\omega\upsi$ published in 2004~\cite{Cronin-Hennessy04}.  Since then,
updated measurements of the previously known dipion transitions have been 
made~\cite{Cronin-Hennessy07,Bhari09,Lees11a,Lees11b}, and, 
additionally, a wide array of new transitions have been investigated by CLEO,
Belle and \babar: $\chibpj\to\pi\pi\chibj$~\cite{Cawlfield06,Lees11a}, 
$\upsii\goesto\eta\upsi$~\cite{He08,Lees11b,Tamponi12}, 
and both $\pi\pi$ and $\eta$ 
transitions from states above the open beauty threshold,
$\upsivv$ to states below threshold 
($\upsott$)~\cite{Aubert08a,Sokolov09,Chen08}. 

Measurement of the relative rates for various hadronic transitions, together
with invariant mass and angular distributions, particularly for those transitions
mediated by different terms in the multipole expansion, can be particularly
enlightening for our understanding of hadronization.  The past decade has
seen substantial advances in the study of hadronic transitions in the 
bottomonium spectrum as the following discussion shows.

\subsubsection{Progress In $\upsns\goesto\pi\pi\upsms$}

Among the more interesting questions concerning dipion
transitions between $\upsns$ states below open beauty threshold
arose from the stark contrast 
between the invariant mass distributions of the pion pair for the transition 
$\upsiii\goesto\pi\pi\upsi$ as compared to that for $\upsii\goesto\pi\pi\upsi$.  
The $\pi\pi$ invariant mass distribution in the latter transition agrees well
with that from the analogous 
transition $\psi(2S)\goesto\pi\pi\jpsi$, while the former transition 
has a double-humped invariant mass structure.  While the $2S\to 1S$
transitions can be explained by a fairly simple matrix element, the
$\upsiii\to\upsi$ distribution cannot.  

The CLEO Collaboration studied the decay dynamics of
$(\upstht)\goesto\pi\pi(\upsto)$ using a sample of 
$4.98\esi~\upsiii$~\cite{Cronin-Hennessy07}. The size of this data sample 
enabled at last a detailed study of the decay dynamics, which
was done according to the formalism of Brown and Cahn~\cite{Brown75} and 
the multipole expansion model cited previously.  In addition, the 
branching fractions for all $\upsns\goesto\pi\pi\upsms$ have been updated, 
first by CLEO~\cite{Bhari09}, and more recently 
by \babar~\cite{Lees11a,Lees11b}. Two of these branching fractions, 
 ${\cal{B}}(\upstth)\goesto\pi\pi\upsi$, are now known to a relative uncertainty 
 of less than $1.5\%$~\cite{PDG} and therefore can be more profitably used as 
tagging modes for searches for unusual decays of $\upsi$ (see 
Section~\ref{sec:newphys}).

\subsubsection{$\chibpj\goesto\omega\upsi$}

As noted previously, the observation by the CLEO Collaboration of 
$\chibpj\goesto\omega\upsi (J=1,2)$~\cite{Cronin-Hennessy04} 
represented the first observation of any non-$\pi\pi$ hadronic 
transition between bottomonium states, and also the first
observation of $\chibpj$ in something other than an E1 radiative transition.    
The analysis involved a full reconstruction of the decay chain
$\upsiii\goesto\gamma\chibpj$; $\chibpj\goesto\omega\upsi$; $\upsi\goesto\dilep
(\ell\equiv e,\mu)$ using the full $\upsiii$ CLEO data sample
of $5.9\esi\upsiii$ decays.  The resulting nearly background-free analysis 
revealed branching fractions of $(1.63\asy{0.35}{0.31}~\asy{0.16}{0.15})\%$ 
and $(1.10\asy{0.32}{0.28}~\asy{0.11}{0.10})\%$ for the $\chibpone$ and 
$\chibptwo$ transitions, respectively.  

The $\chibpj\goesto\omega\upsi$ process is expected to proceed via the emission
of three E1 gluons~\cite{Voloshin03}, since the quantum numbers of the $\omega$
are that of the photon, $\jpc = \omm$ and the ratio of rates for the
$\chibptwo$ transition to that for the $\chibpone$ transition is expected to be
$1.3\pm 0.3$.  The measured branching fractions for the $\omega$ transitions
from CLEO yield a ratio of rates between 0.9 and 1.5, in agreement with
this prediction. 

\subsubsection{$\chibpj\goesto\pi\pi\chibjone$} 

Dipion transitions are expected to be the dominant transitions between radially
excited heavy quarkonium states and their lower-mass partners, as they appear
in the lowest possible order in the multipole expansion~\cite{Yan80}.
Transitions from the $\chibpj$ and $\chibj$ states of the same $J$ (and in fact
the states $\etabp$ and $\etab$) have been long expected, and in 2006, CLEO
published the first observation of $\chibpj\to\pi\pi\chibj$ transitions. 

The CLEO analysis \cite{Cawlfield06} 
utilized the involved a full reconstruction of the decay chain
$\upsiii\to\gamma\chibpj$; $\chibpj\to\pi\pi\chibj$; $\chibj\to\gamma\upsi$; 
$\upsi\to\dilep$ and again used the full CLEO $\upsiii$ data set.  The results
from both $\pipi$ and $\piz\piz$ analyses were combined, and
branching fractions of $(6.0\pm 1.6\pm 1.4)\%$ and $(8.6\pm 2.3\pm 2.1)\%$,
for $J=2$ and $J=1$, respectively~\cite{PDG}.

A study by \babar published in 2011 \cite{Lees11a}, using a much larger sample
of $\upsiii$ decays, obtained precise measurements of the branching
fractions: $\br(\chibpone\goesto\pipi\chibone) = (9.2\pm 0.6 \pm 0.9)\%$
and $\br(\chibptwo\goesto\pipi\chibtwo) = (4.9\pm 0.4 \pm 0.6)\%$.

\subsubsection{$\upsii\goesto\eta\upsi$ and search for related transitions}

The decay $\psi(2S) \to \eta J/\psi$ was one of the first hadronic transitions
observed in charmonium, with a branching fraction of ${\cal B}[\psi(2S) \to
\eta J/\psi] = (3.28 \pm 0.07)\%$\cite{PDG}, which is quite large given the
tiny available phase space.   By contrast, until a few years ago, only an upper
limit was known for the corresponding bottomonium process:
${\cal B}[\upsii\goesto\eta\upsi] < 2 \times 10^{-3}$ \cite{PDG06}.

The production of a pseudoscalar meson in hadronic transitions between
two $\tso$ states involves the flip of a heavy quark spin
\cite{Yan80,Kuang06,Voloshin08}.  Such transitions in bottomonium thus
can probe the chromomagnetic moment of the $b$ quark.  Their rates are
expected to scale as $\Gamma \propto (p^*)^3/m_Q^4$, where $m_Q$ is the mass
of the heavy quark $Q = c,b$.  Then one predicts
\beq
\frac{\Gamma[\upstth \to \eta \upsi]}{\Gamma[\psi(2S) \to \eta
J/\psi]} = (0.0025,0.0013)~,
\eeq
implying ${\cal B}[\upstth \to \eta \upsi] = (7.8,6.4)
\times 10^{-4}$ using the latest values from Ref.\ \cite{PDG}.  An explicit
potential model calculation~\cite{Kuang06} 
obtains similar values:  $(6.9,5.4) \times 10^{-4}$.

The transition $\upsii \to \eta \upsi$ was first observed by the CLEO
Collaboration in about 9 million $\upsii$ \cite{He08}.  The rate was about
a factor of (four,three) below the (scaling,potential) prediction.  In just
under 100 million $\upsii$ decays, 
the \babar Collaboration essentially confirmed this
result \cite{Lees11b}.  More recently, the Belle Collaboration, in 158 million
$\upsii$, obtained a rate about 50\% higher than CLEO's or {\sc BaBar}'s
\cite{Tamponi12}.  These results are summarized in Table \ref{tab:etapi0}, for
the transitions $\upstth\to(\eta,\piz)\upsot$.  

\begin{table}
\caption{Branching fractions and upper limits (in units of $10^{-4}$)
for hadronic transitions between $\upsns$ levels involving $\eta$
and $\pi^0$ emission.
\label{tab:etapi0}}
\begin{center}
\begin{tabular}{c c c c c} \hline \hline
 & & CLEO \cite{He08} & \babar \cite{Lees11b} & Belle \cite{Tamponi12} \\
$\upsii$ & $N(\Upsilon)$ (10$^6$) & $9.32\pm0.14$ & $98.6\pm0.9$ & $158\pm4$ \\
         & $\to \eta \upsi$ & $2.1^{+0.7}_{-0.6}\pm0.3$ & $2.39\pm0.31\pm0.14$
 & $3.57\pm0.25\pm0.21$ \\
         & $\to \pi^0 \upsi$ & $< 1.8$ & -- & $0.41$ \\
$\upsiii$ & $N(\Upsilon)$ (10$^6$) & $5.88\pm0.10$ & $121.8 \pm 1.2$ & -- \\
         & $\to \eta \upsi$ & $< 1.8$ & $<1.0$ & -- \\
         & $\to \pi^0 \upsi$ & $<0.7$ & -- & -- \\
         & $\to \pi^0 \upsii$ & $< 5.1$ & -- & -- \\ \hline\hline
\end{tabular}
\end{center}
\end{table}

The CLEO and Belle upper limits for the isospin-forbidden transition $\upsii
\to \pi^0$, and the CLEO limit for $\upsiii \to \pi^0 \upsi$, can be compared
with predictions based on scaling from the ratio of $\psi(2S) \to \pi^0 J/\psi$
and $\psi(2S) \to \eta J/\psi$. Updating the calculation in Ref.\ \cite{He08}:
\beq
\frac{{\cal B}[\upstth \to \pi^0 \upsi]}
     {{\cal B}[\upstth \to \eta  \upsi]} = (15.9\pm1.3,0.41\pm
0.03)\%~.
\eeq
The Belle upper limit ${\cal B}[\upsii \to \pi^0 \upsi] < 0.41 \times 10^{-4}$
is in mild conflict with the first of these values.

\subsubsection{$\upsivv\goesto X + \upsns\;[X = \pi\pi, \eta]$\label{sec:ups5shadronic}}

Because the $\upsiv$ and $\upsv$ resonances lie above open-bottom threshold, 
they are much broader than the below-threshold $\upsott$. Their dominant 
decays are to those open-bottom mesons whose pair-production threshold lies 
lower in mass.  
The large $\upsivv$ data samples accumulated by \babar and Belle make
possible, despite their small expected branching fractions, the observation
of hadronic and radiative transitions from $\upsiv$ and $\upsv$ ---
and their observation yields additional interesting and 
important information concerning the nature of the bottomonium system.

Both \babar and Belle have studied the transitions 
$\upsiv\to\pipi\upsot$ using their large $\upsiv$ data samples, exclusively
reconstructing the full transition, with the $\pipi$ observed in conjunction
with the leptonic decay $\dilep=(\ee,\mumu)$ of the daughter $\upsns$ state. 
The partial widths for $\upsiv\to\pipi\upsot$ thus obtained are similar to 
the partial widths for $\upsott\to\pipi\upsot$~\cite{Aubert08a,Sokolov09}.  
This is what one would have na\"{\i}vely expected, though the \babar analysis 
made an important observation concerning the invariant mass of the $\pipi$
pairs emitted in the $\upsiv\to\pipi\upsii$ transition.  That is, the
distribution of the $\pipi$ invariant mass in this transition bears a striking
resemblance to that of the $\upsiii\to\pipi\upsi$ transition.  
Interestingly, both are
$\Delta n = 2$ transitions, while the $\Delta n = 3$ transitions resemble
more closely their $\Delta n = 1$ counterparts.  

By contrast, the transitions $\upsv\to\pipi\upsi$ observed by the Belle 
Collaboration showed significant discrepancies with respect to theoretical 
expectations.  In a sample of $21.7~\invf$ of $\ee$ collisions at an energy
corresponding to $M(\upsv)$, Belle observed very large rates for the 
transitions $\upsv\to\pipi\upsott$: up to 100 times larger than 
the corresponding rates for $\pipi$ transitions among sub-threshold 
$\upsns$ states~\cite{Chen08}.  This was so far above the expectations of 
similar rates
that immediately speculation ensued concerning the possible existence of
tetraquark or other exotic non-$\upsns$ states near 
$\upsv$~\cite{Ali10}.  Table~\ref{ups5stable} summarizes the results of the
rate measurements for $\upsv\to\pipi\upsns$, and includes the rates for 
$\upsott\to\pipi\upsot$ for comparison.

\begin{table}
\caption{Comparison of measurements of $\upsms\to\pipi\upsns$ transitions, 
where $m\in(2,5)$ and $n\in(1,3)$.\label{ups5stable}}
\begin{center}
\begin{tabular}{c c c c} 
\hline \hline
Process & $\Gamma_{total}$ & $\Gamma_{\pipi\upsns}$ & Reference \\
& $\mev$ & $\kev$ & \\
\hline
$\upsii\to\pipi\upsi$ & 0.032 & 6.0 & \cite{PDG} \\
$\upsiii\to\pipi\upsi$ & 0.020 & 0.9 & \cite{PDG} \\
$\upsiii\to\pipi\upsii$ & 0.032 & 0.6 & \cite{PDG} \\
$\upsiv\to\pipi\upsi$ & 20.5 & 1.8 & \cite{PDG} \\
$\upsiv\to\pipi\upsii$ & 20.5 & 1.7 & \cite{PDG} \\
$\upsv\to\pipi\upsi$ & 110& 590 & \cite{Chen08} \\
$\upsv\to\pipi\upsii$ & 110 & 850 & \cite{Chen08} \\
$\upsv\to\pipi\upsiii$ & 110 & 520 & \cite{Chen08} \\
\hline
\hline
\end{tabular}
\end{center}
\end{table}

Given the observation of $\pipi$ transitions from $\upsns$ states 
above open-bottom threshold, it might be expected that other hadronic
transitions from such states might also be observable.  Using their full
$\upsiv$ data sample, the \babar Collaboration has observed the transition
$\upsiv\to\eta\upsi$, with $\eta$ decaying to $\pipi\piz$
and the $\upsi$ decaying leptonically.  The reported branching fraction of 
$(1.96\pm 0.11)\times 10^{-4}$~\cite{Aubert08a} is unexpectedly high ---
approximately 2.5 times greater than the branching fraction for the
dipion transition to $\upsi$.  

\section{PRODUCTION}

The production of heavy quarkonia in hard-scattering processes involves
different momentum scales in both perturbative and nonperturbative 
regimes of QCD. A detailed understanding of this process is therefore 
an important test of our understanding of QCD. 

A number of different approaches have been proposed to factorize the  
high-momentum (short-distance) scale process leading to a $Q\bar Q$ 
pair (predominantly by gluon-gluon fusion diagrams) and the low-momentum 
(long-distance) scale of the process that binds the $Q\bar Q$ into 
color singlet quarkonia of the given quantum numbers: the color singlet 
model (CSM)~\cite{Chang80,Baier81,Baier83}, NRQCD 
factorization~\cite{Bodwin95}, fragmentation function 
factorization~\cite{Nayak05}, and $k_T$ factorization~\cite{Baranov02,Kniel06}. 

The mutual relations and differences among these approaches are discussed 
in detail in~\cite{Brambilla11}. One of the open issues regards the 
contribution of $Q\bar Q$ pairs that are in color singlet (CS) 
as compared to those in which the 
$Q\bar Q$ pairs are in a color octet (CO). In the QCD-based CSM 
only the former is considered, while
in the NRQCD, $k_t$ and fragmentation approaches 
both singlet and octet contributions exist.

The parameters relevant for the calculation of the color singlet contribution
can be extracted from measurements of quarkonium decays using potential models
or lattice calculations, so the inclusive differential cross section has no 
additional free parameters. It is not possible at present to relate the 
parameters of the color-octet contribution to quantities measured in 
quarkonium decays, and they are usually determined from fits to 
differential cross sections. 

Bottomonium is heavier and less relativistic than charmonium, so the 
agreement between theory and experiment is expected to be better for 
bottomonia.
However the cross sections for bottomonia are much smaller, and until the
advent of the LHC, most of the measurements of heavy quarkonium hadroproduction 
involved $J/\psi$ and $\psi(2S)$.
Also, experimentally it is more difficult to determine the $\upsns$ that 
are directly produced subtracting the contribution originating from the 
decays of higher resonances, since the spectrum below the open-flavor
threshold is much richer compared to that of charmonium.

\subsection{Differential cross sections for $\upsns$ vs.\ theory}

One of the key features of the differential cross section for direct 
$\upsns$ production is its $p_t$ dependence. At leading order the 
color singlet contribution has a $p_t^{-8}$ dependence, while the 
color octet contribution has a $p_t^{-4}$ dependence due to gluon 
fragmentation.  However at NLO the color singlet contribution has 
large corrections proportional to $p_t^{-6}$, and a small correction 
proportional to $p_t^{-4}$ due to quark fragmentation. The NNLO 
corrections have been only partially calculated, but the gluon 
fragmentation provides large corrections proportional to $p_t^{-4}$.

No significant rapidity ($y)$ dependence is expected in the
central region. In the high $p_t$ limit the shapes of the 
differential cross sections are predicted to be the same for all the $\upsns$.

For the color singlet contribution the ratios of the $\upsns$/$\upsi$
cross sections are simply related to the ratios of wavefunctions at the origin.

The CDF experiment measured the inclusive differential cross 
section, as a function of $p_t$, for $\upsi$, $\upsii$ and 
$\upsiii$ production in $p\bar p$ collisions at $\sqrt{s}=1.8\,$TeV 
in the rapidity interval $\vert y\vert< 0.4$~\cite{Acosta02}. The 
D0 experiment measured the inclusive differential cross sections for
$\upsi$ and $\upsii$ in a broader interval,
$\vert y\vert< 1.8$~\cite{Abazov08}. 

The $\upsi$, $\upsii$, and $\upsiii$ differential cross sections 
have been measured at LHC in $p p$ collisions at $\sqrt{s}=7\,$TeV. 

The CMS experiment has measured the differential cross sections 
as a function of $p_t$ for $p_t<30\,$GeV/c$^2$ in the rapidity 
range $\vert y\vert <2 $~\cite{Khachatryan11}. They find that 
the $p_t$ dependence of the three cross sections is in excellent 
agreement with the Tevatron measurements despite the substantial differences 
in collision energy and $y$ interval. 

The LHCb experiment, whose detectors cover the forward rapidity 
region, has presented measurements of the double differential 
cross section as a function of $p_t$ and $y$ for 
$p_t<15\,$GeV/c$^2$ and $2.0<y<4.5$~\cite{Aaji12a}.

Finally, the ATLAS experiment has recently extended the measurement of 
these cross sections up to $p_t<70\,$GeV/c$^2$ in the rapidity 
range $\vert y\vert <2.25$  based on an integrated luminosity 
of $1.8~\invf$~\cite{Aad12p}. The measured cross sections 
are in agreement with the measurements from CMS and LHCb, in 
the rapidity ranges covered by both experiments.

The measured cross sections are qualitatively in agreement with the various
predictions, but ultimately no prediction is able to describe cross sections
accurately in all $p_t$ ranges. The comparison is 
made more difficult by the uncertainties in the $\upsns$
polarization, which result in uncertainties in acceptance 
corrections, and by the uncertainties in the contribution to 
the inclusive differential cross sections from decays of higher resonances.

The ratios of the $\upstth$ to $\upsi$ cross 
sections as a function of $p_t$ have been measured by 
CMS~\cite{Khachatryan11} and ATLAS~\cite{Aad12p}.
Both experiments observe an increase for both ratios up 
to $p_t\approx 30\,$GeV/c$^2$. At larger $p_t$, so far 
measured only by ATLAS, the cross section ratios seem to 
reach a saturation. At large $p_t$ the direct production 
is expected to dominate over indirect contributions from 
decays of higher resonances, and ATLAS observes that the 
values of the plateau are somewhat large, but compatible, 
with the values of the ratios of the wavefunction at the 
origin as expected for the color singlet contribution.

\subsection{Polarization of $\upsns$}
The angular distribution of the two leptons from the decay 
of a vector state can be written as a function of the angles 
of the outgoing leptons with respect to a given frame where 
the polar angle $\theta$ is along the quantization axis,
$W(\theta,\phi)\propto (1+\lambda_\theta\cos\theta 
+\lambda_\phi\sin^2\theta\cos{2\phi}
+\lambda_{\theta\phi}\sin{2\theta}\cos\phi)$ and the 
parameter $\lambda_\theta$ is 0 or 1 for 100\% longitudinal 
or transverse polarization. 

The reference frames adopted in the literature are more than 
one: the most used are the Collins-Soper  frame, where the 
quantization axis is chosen as the bisector of the beam 
directions in the $\Upsilon$ frame, and the helicity frame, 
where the quantization axis is chosen along the $\Upsilon$ 
direction in the collision center of mass frame. It must be 
noted that the two definitions are not 
equivalent~\cite{Faccioli12}. It is also possible to 
express the polarization in terms of frame-independent 
quantities~\cite{Faccioli10}.

The $\upsns$ from color singlet are expected to be produced 
with longitudinal polarization \cite{Artoisenet08}, while 
NRQCD, where color octet contributions are significant, predicts 
a strong transverse polarization for quarkonia produced at 
high $p_t$\cite{Gong11}. 

The CDF \cite{Acosta02,Aaltonen12} and D0 experiments~\cite{Abazov08}
measured the $\upsi$ polarization in $p\bar p$ collisions at $\sqrt{s}=1.8\,
$TeV, obtaining results that do not agree between the two experiments 
and are also in disagreement with NRQCD predictions.

The CMS experiment has recently presented a new measurement of $\upsiii$,
$\upsii$ and $\upsi$ angular distribution at $\sqrt{s}=7\,$TeV in three
different quantization frames~\cite{Chatrchyan12}, finding no
evidence for either large transverse or large longitudinal polarization.

\subsection{How much hadronic production proceeds via P-wave states?}

The comparison between measurement and theoretical prediction 
for the total and differential cross sections of
any of the bottomonium states requires the knowledge of the 
fraction of events that originate from the decay of higher 
bottomonium resonances. This knowledge becomes crucial when 
measuring the $\upsns$ polarization, because the angular 
distribution of $\upsns$ from $\chibj$ decays is significantly different.

The first determination of the fraction of $\upsi$ originating
from $\chibjone$ in $p\bar p$ collisions at $\sqrt{s}=1.8\,$TeV 
was presented by CDF \cite{Affolder00}: for $p_t> 8\,$ GeV/c$^2$ 
the $\chibj$ and $\chibpj$ decays account for
$27.1\pm6.9\pm4.4\%$ and $10.5\pm4.4\pm1.4\%$ respectively of the 
observed $\upsi$ yield. The fraction of directly produced $\upsi$, 
accounting also for the feed-down from $\upsii$ and $\upsiii$, 
is $50.9\pm8.2\pm9.0\%$.

A new measurement has recently been presented by LHCb \cite{Aaij12}
in $p p$ collisions at $\sqrt{s}=7\,$TeV: for $6<p_t<15\,$GeV/c$^2$
the fraction of $\upsi$ originating from $\chibj$ is on average $20.7 \pm5.7
\pm2.1^{+2.7}_{-5.4}\%$.  They do not observe a dependence of this fraction
on $p_T$ in the range studied.

\section{NEW PHYSICS ASPECTS~\label{sec:newphys}}

Searches for physics beyond the standard model are often the 
domain of experiments that probe the highest available energies, or, at lower 
energies, involve the decays of open-flavored heavy mesons like the $D$, 
$D_s$, $B$ or $B_s$.   Heavy quarkonia, however, also may be used to search for
hints of new physics.  We describe in this section some recent new-physics
searches in studies of bottomonium decays. 

\subsection{Decays of $\upsi\goesto\invis$ via $\upsns\goesto\upsi\pipi$}

Decays of $\upsns$ states to final states that are not observable by typical 
multipurpose detectors offer an excellent place to search for the impact of
physics beyond the standard model, chiefly because the only allowed
``invisible" decays of $\upsns$ within the standard model are the decays
$\upsns\goesto\nu\nubar$, whose branching
fractions can be fairly tightly constrained by the well-measured leptonic
decays of $\upsns$.  The SM prediction for 
${\cal{B}}(\upsi\goesto\invis)$, based on the
present measurement of the leptonic decay rate of $\upsi$, is $\approx 1 \times
10^{-5}$~\cite{Chang98}.  
If the rate of invisible decays of $\upsns$ is observed to be substantially 
larger than the predicted rate, the presence of additional new mediating
bosons, which would increase the expected $\upsns\goesto\nu\nubar$ rate, 
could be the cause.  Another potential explanation would involve the decay 
of $\upsns$ to pairs of dark matter candidate particles ---
such an enhancement could increase the 
invisible branching fraction up to near $10^{-3}$~\cite{McElrath05}.

Recent searches at Belle, CLEO and and {\sc BaBar} for $\upsns$ to invisible 
final states have all used a tagging method in which 
data from collisions at higher $\upsns$ states are used, and a charged 
dipion transition to $\upsi$ is observed.  In general, the study of these 
invisible $\upsns$ decays
depends on the ability of the detector in question to trigger on very
low-multiplicity events, and the suppression of ``normal'' $\upsi$ decays to
final states which should in principle be detectable. 

The first of these searches was published by the Belle Collaboration
using $\pipi$-tagged $\upsiii\goesto\pipi\upsi$ decays
taken from a sample of $11\esi~\upsiii$ decays, 
and yielded a $90\%$ upper limit for the branching fraction 
${\cal{B}}(\upsi\goesto\invis)$ of 0.25\%~\cite{Tajima07}.  A
similar analysis by the CLEO Collaboration using their sample of $8.7\esi~\upsii$ 
decays yielded a somewhat larger upper limit of $0.39\%$~\cite{Rubin07}.  The 
best upper limit for this branching fraction comes from the {\sc BaBar} 
Collaboration, and uses their very large data sample of $91\times 10^6~\upsiii$
decays, lowering the upper limit by an order of magnitude, to $3.0\times
10^{-4}$ at 90\% confidence~\cite{Aubert09a}.  Despite the improvement, 
this limit is still somewhat more than an order of magnitude above
the SM prediction.   

\subsection{Search for $\upsi\goesto\gamma + \invis$}

In addition to the searches for purely invisible final states arising from
$\upsi$ decay, the \babar~Collaboration 
published a study of $\upsi$ decays to a single photon plus a 
particle which decays to a two-body invisible final 
state, $\upsi\goesto\gamma\lmh;\lmh\goesto\chi\chi$~\cite{Sanchez11}.  
The $\upsi$ in this study are tagged in a similar manner to those
discussed above, although in this case the analysis used a sample of
$98.3\esi$ events.  The final state is then tagged by the observation of a
single photon.  The search obtained no significant yield of events above the
expected background, yielding 90\% confidence 
upper limits on the product branching fraction for scalar $\lmh$:
$\br(\upsi\goesto\gamma\lmh)\times\br(\lmh\goesto\invis)$ 
$ < 1.9-37 \emsi$ for $M(\lmh) < 9.2~\gevm$.  

\subsection{Decays $\upsns\goesto\tau^{+}\tau^{-}$ and tests of lepton
universality}

The $\upsns$ states can decay by $b \bar b$ annihilation to a virtual
photon which then will materialize to anything kinematically permitted,
e.g., the lepton pairs $e^+e^-$, $\mu^+ \mu^-$, and $\tau^+ \tau^-$.  Lepton
universality predicts equal rates for these three processes; phase space
corrections are small.  Scalar or pseudoscalar intermediate states in the
direct channel, such as light Higgs bosons, will modify these predictions,
as their couplings to the leptons will be proportional to the lepton mass.
Possible mixing of the $\etabn$ states with light mass Higgs can also impact
the relative branching fractions of $\upsns\to\dilep$,
since for $M(\lmh)$ above
about 9.4 $\gevm$, the dominant leptonic 
decay mode of $\lmh$ is expected to be $\tautau$, and $\lmh$ may mediate
the process $\upsi\goesto\gamma\etab;\;\etab\goesto\lmh\goesto\tautau$~\cite{Domingo09,Sanchez10}.  

CLEO measured the ratios ${\cal R}^{nS}_{\tau\tau} \equiv {\cal B}
(\upsns \to \tau \tau)/{\cal B}(\upsns \to \mu \mu)$ for the $n=1,2,3$ states 
with the results shown in Table \ref{tab:nSll} \cite{Besson07}.
No deviation was seen from lepton universality at the few percent level.

\begin{table}
\caption{Ratios ${\cal R}^{nS}_{\tau\tau}$ of $\upsns$ branching
fractions to $\tau^+ \tau^-$ and $\mu^+ \mu^-$ and inferred branching fractions
of $\upsns$ to $\tau^+ \tau^-$ \cite{Besson07,Sanchez10}.
\label{tab:nSll}}
\begin{center}
\begin{tabular}{c c c c} \hline \hline
 & ${\cal R}^{nS}_{\tau\tau}$ &
 ${\cal B}(\upsns \to \tau^+ \tau^-)$ (\%) & Reference \\ \hline
$\upsi$ & $1.02\pm0.02\pm0.05$ & $2.54\pm0.04\pm0.12$ & \cite{Besson07}\\
$\upsi$ & $1.005\pm 0.013\pm 0.022$ & $2.49\pm 0.03 \pm 0.07$ & \cite{Sanchez10} \\
$\upsii$ & $1.04\pm0.04\pm0.05$ & $2.11\pm0.07\pm0.13$ & \cite{Besson07}\\
$\upsiii$ & $1.08\pm0.08\pm0.05$ & $2.52\pm0.19\pm0.15$ & \cite{Sanchez10} \\ \hline \hline
\end{tabular}
\end{center}
\end{table}

The \babar Collaboration has searched for
violations of lepton universality in $\upsi$ decay using their data sample
of $121\esi~\upsiii$ decays~\cite{Sanchez10}.
The analysis determines the ratio of branching fractions, 
$\frac{\br(\upsi\goesto\tautau)}{\br(\upsi\goesto\mumu)}$
in the $\upsi$ sample provided by
$\upsiii\goesto\pipi\upsi$ decays.  
One may ``tag'' the presence of the $\upsi$ in these transitions 
without reconstructing by using four-momentum conservation.  That is, if the
$\pipi$ are recoiling against a daughter $\upsi$, the recoil mass $M^2\rec
\equiv (M^2\recoil(\pipi) = s + M_{\pip\pim}^2 - 2\sqrt{s} E^{*}_{\pip\pim}$
will peak at $M(\upsi)$. By selecting $M\rec$ near $M(\upsi)$, the 
other final-state particles, which must originate from the decay of $\upsi$, 
may then be studied.

In this analysis, a final state containing exactly four charged tracks was
required:  $\upsi\goesto\mumu$ was identified by requiring positive muon
identification for the tracks 
and a total reconstructed energy consistent with the initial state;  the decay
$\upsi\goesto\tautau$ was identified using only single-prong decays of $\tau$.
From the yields, trigger, analysis and reconstruction efficiencies,
the ratio $\frac{\br(\upsi\goesto\mumu)}
{\br(\upsi\goesto\tautau)} = 1.005\pm 0.013\pm 0.022$ was obtained.
Hence no significant deviation of this ratio from the expected SM value 
of 0.992 was observed.  
This marks a substantial improvement in precision with respect to the 
previous CLEO measurement, and excludes an $\lmh$ with mass lower than 
9 $\gevm$ at the 90\% confidence level.

\subsection{Search for charged lepton flavor violation in $\upsns$ decays}

Another venue for the possible observation of new physics in bottomonium 
decay is the decay of vector bottomonia to lepton pairs that violate lepton 
flavor (e.g., $\upsns\goesto\emu$).  While lepton flavor 
violation is now well established in the neutrino sector due to oscillations 
between neutrino flavors, it is generally agreed that 
lepton flavor violation in the charged sector (CLFV) is so highly 
suppressed that its
observation at any level is an unambiguous sign of new physics.  A number of 
efforts to observe CLFV have been made in $B$ and $K$ decays, or in decays 
of leptons themselves --- but the study of charged lepton flavor violating 
decays of $\upsns$ has only recently been explored in any 
detail.  Both CLEO~\cite{Love08} (using data at $\upsns$ with $n=1,2,3$) 
and \babar~\cite{Lees10} (only $n=2$ and $3$) have studied decays of $\upsns$ 
that violate lepton flavor.  In the case of the CLEO analysis, the $\tau$ lepton 
was identified in its decay to $\nu_{\tau}\nubar_{e} e$ while the \babar~analysis 
used both leptonic and several hadronic $\tau$ decays.   90\% CL upper limits 
for the branching fractions for CLFV modes of all $\upsns$ states are below
$10\emsi$, implying lower limits for the mass associated with CLFV 
operators of order 1 TeV~\cite{Lees10}.

\subsection{Searches for low-mass Higgs in $\upsns$ decays}

Frank Wilczek wrote one of the first papers concerning the appearance of 
new physics in heavy quarkonium decays~\cite{Wilczek77} soon after the 
the announcement of the first observations of bottomonium at Fermilab, 
in which he calculated the ratio of the rate of the decay of heavy 
vectors V to Higgs + $\gamma$ to the rate to $\mumu$: 
\begin{equation}
\frac{\Gamma(V\goesto H\gamma)}{\Gamma(V\goesto\mumu)}= 
\frac{G_Fm_q^2}{\sqrt{2}\pi\alpha}\left(1-\frac{m_H^2}{m_V^2}\right)^{1/2}.
\end{equation}

Under several extensions of the standard model, including the next-to-minimal
supersymmetric standard model 
(NMSSM)~\cite{Hiller04,Dermisek05,Dermisek07,Kim84}, 
the existence of light Higgs bosons is postulated. In particular 
the NMSSM can have a CP-odd neutral Higgs boson with mass less
than 10 $\gevm$, making it observable in radiative decays 
$\upsns\goesto\gamma\lmh$.   Predictions for the branching fractions for
such decays are as large as $10^{-4}$~\cite{Dermisek05}. 
The most recent searches for CP-odd light Higgs bosons in $\upsns$ radiative
decay have been performed at CLEO and \babar.  

CLEO undertook a study of $\upsi\goesto\gamma+(\mumu,\tautau)$ based on 
$21.5\efo\upsi$ decays, in which full reconstruction of the $\gamma\mumu$ 
final state was attempted, while $\gamma\tautau$ events were selected based on 
events having a large missing energy and at least one well-identified $\mu$ 
or $e$ in addition to a well-reconstructed $\gamma$.  The 90\% confidence
limits on the product branching fraction 
$\br(\upsi\goesto\gamma\lmh;\lmh\goesto\dilep)$ for $\dilep=(\mumu,\tautau)$ 
were then calculated to range from $(1-20)\emsi~((1-48)\emfi)$ for
$\lmh\goesto\mumu~(\tautau)$ with $\lmh$ masses in the range $M(\lmh)<3.6~\gevm$
($2M_{\tau} < M(\lmh) < 9.5~\gevm$).  The $\tautau$ results improved
previous limits from ARGUS by more than two orders of magnitude, 
and remain the most stringent constraints on NMSSM models from
this $\lmh$ decay mode.  

\babar has searched for light Higgs bosons in several visible decay 
modes~\cite{Lees11c,Lees12,Aubert09b,Aubert09c} in addition to the 
searches in invisible modes previously discussed~\cite{Sanchez11}.  
The searches for $\lmh\goesto\mumu$ at $\upsns$ with $n=1,2,3$ 
required full reconstruction of the final state, namely a well reconstructed 
photon in addition to two well-identified muons.  
Using data samples of $99\esi\upsii$ and $122\esi\upsiii$ decays, seeing no 
evidence for the decay of $\lmh\goesto\mumu$ in radiative events, \babar 
reported 90\% upper limits of $\br(\upsns\goesto\lmh;\lmh\goesto\mumu)$ 
of $(0.26 - 8.3)\emsi~((0.27 - 5.5)\emsi)$ for $n=2~(n=3)$, and for masses
of $\lmh$ ranging from $2M_{\mu}$ to 9.3 GeV/$c^2$~\cite{Lees11c}.  \babar also 
searched for $\lmh$ in $\upsi$ radiative
decay using $\pipi$-tagged $\upsi$ mesons in $\upsii(\upsiii)$ data samples 
containing $92.8\esi~(116.8\esi)$ events, and set 90\% confidence 
product branching fraction limits of 
$\br(\upsi\goesto\gamma\lmh;\lmh\goesto\mumu)$ of $(0.28- 9.7)\emsi$
for $\lmh$ with masses ranging from $2M_{\mu}$ to 9.2 GeV/$c^2$~\cite{Lees12}.

\babar also performed searches for $\lmh$ in radiative $\upsiii$ decays in 
which the $\lmh$ decays to $\tautau$, with  
both $\tau$ required to decay leptonically.  The event selection required 
events to have exactly two charged tracks, identified as
either $e$ or $\mu$, and a signal photon with energy greater than 
$100\mev$. The range of 
$\lmh$ masses tested was $(4.03-10.10)~\gevm$, with a small range of 
exclusion from (9.52-9.61) $\gevm$ to avoid peaking background
due to two-photon cascades from $\upsiii$ through $\chibpj$.  No 
evidence for $\lmh\goesto\tautau$ was observed, and upper limits
of $\br(\upsiii\goesto\gamma\lmh;\lmh\goesto\tautau)$ of $(1.5-16)\emfi$ 
were set.  

\section{CONCLUSIONS}

Since the discovery of the first bottomonium state in 1977, our understanding
of the $b \bar b$ system has made steady progress.  Comparison of the
bottomonium and charmonium spectra provided strong evidence for the
flavor-independence of the strong force.  The bottomonium spectrum was
surprisingly well described by non-relativistic quantum mechanics, with
relativistic embellishments accounting reasonably well (though not perfectly)
for fine and hyperfine structure.  Significant advances were made by lattice
QCD once light-quark degrees of freedom could be taken into account.
Above $B \bar B$ threshold, new states $Z_b(10610)$ and $Z_b(10650)$ were seen
which could be interpreted as $B \bar B^* + {\rm c.c.}$ and $B^* \bar B^*$
molecules.

Transitions of bottomonium have continued to be understood in greater and
greater detail over the years.  Electric dipole transitions were qualitatively
predicted in the non-relativistic approximation, with small corrections serving
to distinguish among relativistic schemes.  Such transitions led to the
discovery of the first D-wave bottomonium state, the $\Upsilon(^3 1 {\rm D}_2)$
(where the spin $J=2$ is the most probable).  Forbidden magnetic dipole
transitions, whose magnitude was difficult to anticipate, provided gateways
to the $\eta_b(1 {\rm S})$, while that state and the $\eta_b(2 {\rm S})$ were
seen in electric dipole transitions from the $h_b(1{\rm P})$ and
$h_b(2{\rm P})$.  In turn, the $h_b(1{\rm P})$ (first discovered in the decay
$\Upsilon(3S) \to \pi^0 h_b(1{\rm P})$) and $h_b(2{\rm P})$ were seen in
the surprisingly strong transitions $\Upsilon(5{\rm S}) \to \pi^+ \pi^-
h_b([1,2]{\rm P})$, via the gateways of $Z(10610)$ and $Z(10650)$.

Information on quarkonium production has helped to understand the subtleties
and limitations of perturbative and nonperturbative QCD.  A key issue, still
unresolved, is the role of color-octet pairs in the $Q \bar Q$ wavefunction.
Descriptions of production cross sections seem to be only marginally improved
with their inclusion, and the jury is still out regarding their role in
hadronic quarkonium decay.  Details still under study include the dependence of
production cross sections on transverse momentum $p_T$ and rapidity $y$, the
polarization of hadronically produced $\Upsilon$ states (found to be small),
and the role of feed-down from hadronically produced $\chi_b$ states,
accounting for several tens of percent of the hadronically produced $\Upsilon$
levels.  

The study of bottomonium production in inclusive hadronic interaction is about
to enter an era of precision measurements: The LHC experiments will have the
opportunity to measure the direct cross sections and polarization,
disentangling the contribution of higher resonances.

Finally, thanks to recent contributions from
CLEO and \babar over the past several years in their
studies of bottomonium decays, additional constraints on 
new physics have been determined.  These experiments have 
searched for hints of new physics in the direct decays of 
$\upsi$ to invisible final states, constraining the overall 
branching fraction to such final states to be less than 
$3\emfo$, and additionally the decays of $\upsi$ to a radiative
photon plus an invisible final state, whereby limits on the 
product branching fractions ${\cal{B}}(\upsi\to\gamma\lmh)\times 
{\cal{B}}(\lmh\to{\rm invisible})$ were set for a wide range of
masses of light mass Higgs candidates $\lmh$.  

Light mass Higgs bosons have also been studied in radiative decays
of $\upsns$ states, where the $\lmh$ decays leptonically.   Relatively 
stringent limits for $\lmh$ between $2 M_\mu$ 
down to as low as few parts in $10^7$ (and up to approximately 
$1\times 10^5$) were set on the product branching fractions 
${\cal{B}}(\upsi\to\gamma\lmh)\times {\cal{B}}(\lmh\to\mumu)$.  Somewhat
higher limits for decays to $\tautau$ were obtained (a few times $10^{-5}$).  

Leptonic decays of $\upsns$ also prove to be a testing ground for
new physics.  CLEO and \babar have done extensive studies of their
large $\upsns$ data sets for violations of lepton universality or 
of lepton number, both of which would require new physics to explain.  

We look forward to many further insights on QCD and on new physics signatures
from the study of the rich bottomonium system, from experiments already
completed ({\sc BaBar}, Belle, CDF, D0) and from those in progress or planned
(LHCb, ATLAS, CMS, Belle II, $\ldots$).  Bottomonium has truly been an ideal
laboratory in which to study known interactions and search for new ones.

\section*{ACKNOWLEDGMENTS}

The work of J.L.R. was supported in part by the U.S. Department of Energy
through Grant No.\ DE-FG02-90ER40560.  The Work of T.K.P. was supported in part
by the National Science Foundation through Grant No.\ PHY-1205843.

\end{document}